\documentclass[preprint,aps,pra]{revtex4}
\usepackage{tabularx}
\usepackage{dcolumn}
\usepackage{graphics}% Include figure files
\usepackage{bm}% bold matha
\usepackage[dvips]{graphicx}
\usepackage{tabularx}
\usepackage{amsmath}
\usepackage{amssymb}
\usepackage{longtable}
\usepackage{verbatim}
\usepackage{float} % to fix the positions of figures and tables
\usepackage{fancyhdr} % for header settings
\usepackage{multirow}
\usepackage{xcolor}
\newcommand{\ket}[1]{\lvert#1\rangle} % Ket
\newcommand{\braopket}[3]{\langle #1 | #2 | #3\rangle} % Matrix Element
\newcommand{\expect}[1]{ \langle #1 \rangle} % Expectation value

\begin{document}
\pagestyle{fancy}
\rhead{\thepage}
\chead{}
\lhead{}
\cfoot{}
\rfoot{}
\lfoot{}

%\title{Study of excited charmed and charmed-strange mesons in relativized quark model}
\title{Study of excited $D$ and $D_s$ mesons in a relativized quark model}
\author{Saba Noor}\affiliation{Centre For High Energy Physics, University of the Punjab, Lahore (54590), Pakistan. }
\author{Faisal Akram}\thanks{F. A. acknowledges HEC grant 20-15728/NRPU/R$\&$D/HEC/2021, Pakistan.}\affiliation{Centre For High Energy Physics, University of the Punjab, Lahore (54590), Pakistan. }
\author{Bilal Masud}\affiliation{Centre For High Energy Physics, University of the Punjab, Lahore (54590), Pakistan. }

\begin{abstract}
We use a modified relativistic quark model to study the properties of excited charmed and charmed strange mesons.
We calculate the masses and wave functions of conventional charmed and charmed strange mesons incorporating both spin and $S$-$D$ mixing effects and fit parameters of the potential model with known experimental states using differential evolution technique.
Using Leading Born-Oppenheimer expansion, we also compute the spectrum and wave functions of first gluonic excited state of charmed and charmed strange mesons. We examine the effects of gluonic excitation on the spectrum of resultant hybrid mesons. By using our calculated spectrum and wave functions, we determine the radiative transitions of the conventional and hybrid open charm mesons. We compare our calculations with experimental data and other works. We expect our results will be beneficial in the detection of the charmed and charmed strange conventional and hybrids mesons.
\end{abstract}
\maketitle

\section{Introduction}\label{intro}
Quantum Chromodynamics (QCD) is experimentally well established theory of strong interaction of quarks and gluons. However, describing rich spectrum of hadrons from the first principle is still challenging due to non-perturbative nature of QCD at low energy. Considering this difficulty, many models and theoretical approaches have been adopted  to calculate hadron properties. The quark model, where applicable, is considered to be a convenient tool. According to the quark model, a conventional meson is a bound state of a quark and an antiquark.
%Experimentally one can identify a particle with the help of its $J^{PC}$ value, where $J$, $P$ and $C$ are the total angular momentum, parity and charge conjugation of a %particle respectively.
The model predicts that a meson can have only certain specific values of $J^{PC}$. The mesons with $J^{PC}$ differing from quark-model values are called exotic mesons.
%According to the quark model, parity and charge conjugation of a conventional meson is $(-1)^{L+1}$ and $(-1)^{L+S}$ respectively, where $L$ and $S$ are orbital and spin %quantum numbers of the meson.
In $1993$ an exotic meson in the light quark sector having $J^{PC}$ $=$ $1^{-+}$, was observed by the VES Collaboration~\cite{VES}. After that, many exotic mesons have been observed. Theoretically, exotic mesons could be glueball, tetra quark or hybrid mesons. It is noted that $J^{PC}$ value of these mesons can be allowed or forbidden by the quark model. Lattice QCD successfully describes these exotics \cite{Juge,MorningStar,Jugehyb,latHyb,HybVdata}. Phenomenological models also provide good description of the properties of exotic mesons \cite{fluxTube,fluxTube1,bagModel,godfreybagModel,constGluon,close,semay,semay1}. In open flavour mesons where quark and antiquark have different flavours, $C$ parity no longer remains a good quantum number. In this case states are identified by their $J^P$ values only.
%Thus it is difficult to identify exotics.
The states having the same $J^P$ value are mixed. Indeed, all $J^P$ states are mixed except $0^+$ and $0^-$. Mixing is either the spin mixing or the $S$-$D$ mixing. We will discuss it in detail in Sec. \ref{mixing}.
%(saba)include two line about n in 11.3.2023
Incorporating mixing, where needed, we study the properties of hybrid as well as conventional mesons in open charm meson sector specifically charmed ($D$) mesons and charmed strange ($D_s$) mesons.

Gluonic field plays a dominant role in binding quarks. In conventional mesons, the gluonic field is in the ground state while for hybrid mesons, it is in the excited state. We are particularly interested in studying the $\Pi_u$ state, where gluonic field is in its first excited state. We use the leading Born-Oppenheimer approximation (LBO) for calculating the spectrum of hybrid mesons. In this approximation, a hybrid meson is treated as a diatomic molecule in which quarks and gluons are considered as nuclei and electron respectively. There are two steps involved in applying LBO approximation. In the first step the static potential energy of gluons is determined by treating quark antiquark spatially fixed, and in the second step the motion of quarks is restored through using this potential in Schr$\mathrm{\ddot{o}}$dinger wave equation. Parity of hybrid meson is $\varepsilon(-1)^{L+\Lambda+1}$, where $\Lambda$ is magnitude of the projection of total angular momentum of gluonic field onto the molecular axis. For $\Pi_u$ potential, in which gluonic field is in its first excited state, $\Lambda=1$ and $\varepsilon=\pm1$ \cite{Juge, MorningStar}.

Experimentally immense data is available in $D$ and $D_s$ meson domain. First two charmed mesons ($D^0$ and $D^\pm$) were discovered in Mark I experiments in 1976 \cite{e1,e2}. Their $J^P$ is $0^-$. $D^0$ and $D^\pm$ have masses $1864.83\pm0.05$ MeV and $1869.58\pm0.09$ MeV respectively. One year later, first charmed strange meson ($D_s^{\pm}$) of $J^P$ $=$ $0^-$, was observed in DASP Collaboration \cite{e3}. Mass of $D_s^{\pm}$ is $1968.34$ $\pm$ $0.07$ MeV \cite{PDG}. In the same experiment $D_s^{*\pm}$ meson  was also discovered having mass $2112.1\pm0.4$ MeV and $J^P=1^-$. Now we have data for more than twenty well recognized states. Information of their masses, $J^P$ values and dozens of their decay modes are available in the Particle Data Group \cite{PDG}. Side by side, a lot of work has been done to calculate their masses and other properties. By using Schr$\mathrm{\ddot{o}}$dinger-like wave equation in relativistic dynamics, spectrum and wave functions of $D$ and $D_s$ mesons are calculated in Ref.~\cite{GI85, godfrey2016, ly, ly2015}. In Ref.~\cite{godfrey2016}, strong and radiative transitions are also reported. Ref.~\cite{diracDs, diracD} implemented Dirac formalism to calculate spectrum, radiative transitions, decay constant and leptonic decay width of $D_s$ and $D$ meson system respectively. In Ref.~\cite{betheSalpeter}, spectra and wave functions of $D$ and $D_s$ mesons are obtained by using Bethe-Salpeter equation. Ref.~\cite{Regge} computed the  spectra of heavy-light meson with the help of QCD motivated relativistic quark model. They investigated Regge trajectories too. By using the Dirac Hamiltonian, Ref.~\cite{DleadingCor} computed the mass spectra, wave functions and hadronic transitions of heavy-light mesons.
They also added leading order corrections in $1/m_{c,b}$. In Ref.~\cite{wlattice}, spectrum of $D$ and $D_s$ mesons is calculated by using Wilson twisted mass lattice QCD.
%(saba)added 4April2023
Ref.~\cite{ly,ly2015} discretized the relativistic wave equation using Cornell potential (Coulomb plus linear potential).
After calculating the masses and wavefunctions of mesons by diagonalizing the resultant Hamiltonian matrix, the spin dependent part of the Hamiltonian is added perturbatively. Since states with different principal quantum numbers can also have same $J^P$. Thus several principal quantum numbers collectively contribute in one mixed state (either spin mixed or $S$-$D$ mixed). In Ref.~\cite{ly,ly2015} the contribution of two to three principal quantum numbers is included to describe a mixed state.
Another limitation of their work is that the perturbative effects of spin dependent potential are included in the masses but not in the wave function. This is understandable because first order perturbative correction to wave function involves a sum over all possible internal states which is extremely difficult to calculate.
We adopted the same approach to solve the relativistic wave equation as did in Ref.~\cite{ly,ly2015} with the improvements that address the limitations of approach of Ref.~\cite{ly,ly2015}. Instead of including spin dependent part of the Hamiltonian perturbatively, we also include spin dependent terms in the Hamiltonian matrix. For this purpose we developed a system of coupled equation for mixed states. Since we are solving the coupled equations nonperturbatively, we are not restricted to include the contribution of only two to three principal quantum numbers to describe a mixed state. 
Two charmed strange mesons named $D^*_{s0}(2317)^{\pm}$ and $D_{s1}(2460)^{\pm}$ attract special attention as the observed values of their masses are much lower than the quark model predictions. Attempts have been made to describe them through alternative models \cite{lat,lat2,DsPuzzle,DsPuzzle1}. 
%lines added from result portion
A lot of work is done in literature to identify the nature of these mesons. In some literature~\cite{tetraquark1,tetraquark2} these states are interpreted as tetraquarks, while in some places $D^*_{s0}(2317)^{\pm}$ and $D_{s1}(2460)^{\pm}$ are taken as $DK$ and $DK^{*}$ molecule states respectively~\cite{dkbarn2003,dkchen2004}. In Refs.~\cite{cc2022,cc2024}, these states are well described by including coupled-channel effects. Ref.~\cite{cc2022} suggests that these states are the mixtures of bare $c\bar{s}$ core and $D^{(*)}K$ component. In Refs.~\cite{godfrey2016, ly, ly2015, chen2020}, these states have been studied as conventional mesons. The exact structure of these particles is still an open question in hadron physics. 
%insert line
%Our results show that the discrepancy significantly reduces when the effect of spin and $S$-$D$ mixing is incorporated.

In this work, we calculate masses, wave functions and radiative transitions of conventional as well as hybrid $D$ and $D_s$ mesons incorporating spin and $S$-$D$ mixing effects on spectrum and wave function. Unlike previous works, we solve Schr$\mathrm{\ddot{o}}$dinger-type equation using full semi-relativistic Hamiltonian with spin-dependent part. This allows us to study both spin and $S$-$D$ mixing without using perturbation theory.

Roadmap of our work is the following:
In Sec.~\ref{potential model}, we describe our potential model for conventional and hybrid mesons. We use an extended version of the famous GI model. For the hybrid mesons, we add the potential model of gluonic excitation which we suggested in our previous work. In this paper we improve the parameter fitting of our proposed gluonic excited potential model by considering the latest lattice data \cite{HybVdata}.
In Sec.~\ref{mixing},  the phenomena of state mixing  are examined. We distinguish two categories of mixed states. We also discuss the effects of spin dependent part of the potential model on the spectrum. In Sec.~\ref{rel wave eq}, we describe our relativistic wave equation. We explicitly describe how we calculate masses and wave functions of mesons using relativistic wave equation. State mixing effects in the Hamiltonian matrix are discussed too. Radiative transitions of charmed and charmed strange mesons are reviewed in Sec.~\ref{radiative transitions}. In the last section we report our results and concluding remarks based on our calculations.

\section{Effective potential model of mesons}\label{potential model}
Phenomenological potential models are found very successful in describing various properties of heavy quark mesons especially mass \cite{GI85,godfrey2016,ly2015,ly,diracDs,diracD,betheSalpeter,cornel,godfreyB,semayPotential,Swanson2005}.
Among the suggested models, the GI model \cite{GI85} is considered to be the most famous one. In this work, we have used it with some modifications.
\subsection{Potential model of conventional mesons}\label{conv.model}
Conventional $D$ and $D_s$ meson is a bound state of a heavy quark ($Q$ = $c$) and a light antiquark ($\bar{q}$ = $\bar{u}$/$\bar{d}$, $\bar{s}$).
Potential model for a conventional meson $V_{Q\bar{q}}(r)$ is
\begin{eqnarray}\label{con.potential}
 V_{Q\bar{q}}(r) &=& -\frac{4\alpha_s(r)}{3r} + br + c + V_{spin-dep}(r),
\end{eqnarray}
where $r$ is inter-quark distance. The first term of the potential model, called Coulomb term, is calculated by one gluon exchange diagram in perturbation theory. It determines the behaviour of potential at small distances. $\alpha_s(r)$ is a strong running coupling constant of QCD. We use its following parameterization \cite{GI85},
%(saba Nov2022) Can we use term parameterization form
\begin{equation}\label{alpha}
  \alpha_s(r) = \sum_{k=1}^3 \alpha_k \frac{2}{\sqrt{\pi}}\int^{\gamma_kr}_0 e^{-x^2}dx,
\end{equation}
where $\alpha_k^{\;,}s$ and $\gamma_k^{\;,}s$ are free parameters.
In this work we take $\gamma_1$ = \large$\frac{1}{2}$\normalsize, $\gamma_2$ = \large$\frac{\sqrt{10}}{2}\;$\normalsize and $\gamma_3$ = \large$\frac{\sqrt{1000}}{2}\;$\normalsize from Ref.~\cite{GI85}. 
%For this choice of $\gamma$'s, $\alpha_2$ and $\alpha_3$ control the shape of $\alpha_s(r)$ function at short distance where it is already fixed by running coupling constant calculate by perturbative loop corrections, whereas $\alpha_1$ controls the shape at large distance. We treat $\alpha_1$ (or equivalently $\alpha_s^{c} \equiv$  \footnotesize$\displaystyle\sum_{k}$\normalsize $\alpha_k$) as a free parameter to be fitted by spectrum, whereas $\alpha_2$ and $\alpha_3$ are fitted by running coupling constant at short distance.
%(changes are made)
With this choice of $\gamma$'s, $\alpha_2$ and $\alpha_3$ control the shape of running coupling at small distance (large $Q^2$), where it is already fixed by pQCD. This allows us to fix $\alpha_2=0.16$ and $\alpha_3=0.20$ by fitting with 2-loop running coupling. Consequently, parameter $\alpha_1$ (or $\alpha_s^c\equiv\alpha_1+\alpha_2+\alpha_3$), which control the shape of running coupling in infrared region, is treated as a free parameter which is to be fixed by spectroscopy ~\cite{sadia,GI85}. Comparison of $\alpha_s(Q^2)$ of the parametrized model with our fitted value of $\alpha_s^c=0.24$ and pQCD 2-loop result is shown in Fig.~\ref{alpha}.
\begin{figure}%[ht]
  \centering
  \includegraphics[scale=0.9]{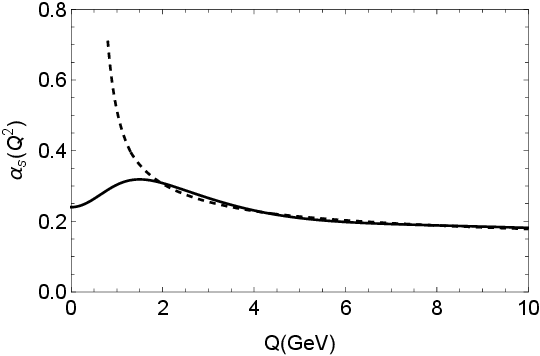}
  %\caption{Comparison of $\alpha_s(Q)$ (parametize model(solid curve) and QCD perturbative calculations(doted curve)).}\label{alpha}
\caption{Strong running coupling constant in momentum space. Solid curve represent data taken from parametrize model and dotted curve represent data from QCD perturbative calculations up to two loop corrections.}
\end{figure}

The second term of the potential is the confinement term dominating at large distances and $b$ is the string tension, determined by fitting the spectrum. The third term $c$ of the potential model is also taken as a free parameter. The last part of the potential model, the spin dependent term $V_{spin-dep}(r)$, is defined as following
\begin{equation}\label{eVsd}
 V_{spin-dep}(r) =  V_{hyp}(r) + V_T(r) + V_{so}(r),
\end{equation}
where $V_{hyp}(r)$ is spin-spin contact hyperfine interaction \cite{Swanson2005}. It is defined as
\begin{equation}\label{Vhyp}
  V_{hyp}(r) = \frac{32\pi}{9m_Q\tilde{m}_{q1}}\alpha_s(r)\left(\frac{\sigma}{\sqrt{\pi}}\right)^3e^{-\sigma^2r^2}\mathbf{S_Q}\cdot\mathbf{S_{\bar{q}}},
\end{equation}
where $\expect{\mathbf{S_Q}\cdot\mathbf{S_{\bar{q}}}}$ = \large $\frac{S(S+1)}{2}-\frac{3}{4}$\normalsize. In Ref.~\cite{GI85} parameter $\sigma$ is
\begin{equation*}
\sigma=\sqrt{\sigma_0^2\left(\frac{1}{2}+\frac{1}{2}\left(\frac{4m_Qm_{\bar{q}}}{(m_Q+m_{\bar{q}})^2}\right)^4\right)+s_0^2\left(\frac{2m_Qm_{\bar{q}}}{m_Q+m_{\bar{q}}}\right)^2}.
\end{equation*}
Numerical values of parameters $\sigma_0$ and $s_0$ are determined by fitting the spectrum. In Eq.~\eqref{eVsd}, $V_T(r)$ is tensor term defined as
\begin{equation}\label{Vtensor}
  V_{T}(r) = \frac{4 \alpha_s(r)}{m_Q\tilde{m}_{q2}}\frac{T}{r^3},
\end{equation}
where $T$ is tensor operator,
\begin{equation}\label{Vhyp}
  T = \mathbf{S_Q}\cdot\hat{r}\mathbf{S_{\bar{q}}}\cdot\hat{r}-\frac{1}{3}\mathbf{S_Q}\cdot\mathbf{S_{\bar{q}}}.
\end{equation}
The tensor operator has nonvanishing matrix elements only between $L>0$ and spin triplet states. Diagonal Hamiltonian matrix entries of tensor operator are \cite{Swanson2005}
\begin{equation}
 \braopket{^3L_J}{T}{^3L_J} = \left\{
      \begin{array}{ll}
       -\frac{L}{6(2L+3)}&\hspace{0.4cm} J=L+1 \\
       \frac{1}{6}&\hspace{0.4cm} J=L \\
        -\frac{(L+1)}{6(2L-1)}&\hspace{0.4cm} J=L-1.
      \end{array}
    \right.
\end{equation}
 To study $S$-$D$ mixed states we also need following off diagonal matrix element of the tensor operator
  \begin{equation}
 \braopket{^3L_J}{T}{^3L'_J} = \frac{\sqrt{(L+1)(L+2)}}{2(2L+3)},
\end{equation}
where $L^{'}$ = $L$ + 2.
In Eq.~\eqref{eVsd}, $V_{so}(r)$ is spin-orbit interaction term. It consists of two parts, such that
\begin{equation}\label{eVso}
V_{so}(r)=V_{sov}(r)+V_{sos}(r).
\end{equation}
Here $V_{sov}(r)$ is spin-orbit vector part
\begin{equation}\label{eVsov}
V_{sov}(r)=\frac{4}{3}\frac{\alpha_s(r)}{r^3}\left[\left(\frac{1}{2m_Q^2}+\frac{1}{m_Q\tilde{m}_{q3}}\right)\mathbf{L}\cdot\mathbf{S_Q}
+\left(\frac{1}{2\tilde{m}_{q3}^2}+\frac{1}{m_Q\tilde{m}_{q3}}\right)\mathbf{L}\cdot\mathbf{S_{\bar{q}}}\right],
\end{equation}
 and $V_{sos}(r)$ is spin-orbit scalar part of spin-orbit interaction term
\begin{equation}\label{eVsos}
V_{sos}(r)=-\frac{b}{r}\left(\frac{\mathbf{L}\cdot\mathbf{S_Q}}{2m_Q^2}+\frac{\mathbf{L}\cdot\mathbf{S_{\bar{q}}}}{2\tilde{m}_{q4}^2}\right).
\end{equation}
For diagonal Hamiltonian matrices, $\expect{\mathbf{L}\cdot\mathbf{S_Q}}$ =
$\expect{\mathbf{L}\cdot\mathbf{S_{\bar{q}}}}$ = \large$\frac{\expect{\mathbf{L}\cdot\mathbf{S}}}{2}$\normalsize, where $\expect{\mathbf{L\cdot S}}$ is
\begin{equation}
\braopket{^1L_L}{\mathbf{L\cdot S}}{^1L_L} = \braopket{^3L_L}{\mathbf{L\cdot S}}{^3L_L} = \frac{J(J+1)-L(L+1)-S(S+1)}{2}.
\end{equation}
To study spin mixed states, we need following matrix element of the $\expect{\mathbf{L.S_Q}}$ and $\expect{\mathbf{L.S_{\bar{q}}}}$ operators:
\begin{align}
\braopket{^1L_L}{\mathbf{L\cdot S_Q}}{^3L_L} = -\frac{\sqrt{L(L+1)}}{2},\\
\braopket{^1L_L}{\mathbf{L\cdot S_{\bar{q}}}}{^3L_L} = \frac{\sqrt{L(L+1)}}{2}.
\end{align}

The form of the spin dependent part of the potential model is extracted from the leading order perturbation theory in the non-relativistic approximation. However, this is not good approximation, especially for light quark in heavy light meson. Due to the presence of light quark in charmed and charmed strange meson, it is intimately important to incorporate relativistic effects. These effects modify the potential in two different ways~\cite{GI85}:
First, quark antiquark separation ($r$) become smeared over a distance of order of inverse quark mass.
 By introducing the smearing function~\cite{GI85}, smearing over $1/r^3$ terms in spin dependent part of the potential model is incorporated.
 Smearing also eliminates the instability in the solution of the relativistic wave equation which occurs at short distances due to $1/r^3$ term in spin dependent potential model, and makes the solution stable even at short distances.
 The second way to incorporate relativistic effects is to make potential model momentum dependent. Momentum dependence in a spin dependent term can effectively be incorporated by introducing multiple constant factors. This is done only for light quark, for heavy quark momenta drop out. Since momentum dependence modifies each spin interaction term in a different way, so we are required to take multiple constant ($\epsilon_i$) interaction dependent, where $i=1$, $2$, $3$ and $4$ is for hyperfine term, tensor term, spin orbit vector term and spin orbit scaler term respectively. We combine these constants into light mass and define interaction dependent mass as~\cite{ly,ly2015}, $\tilde{m}_{qi}$ = $\tilde{m}_{q}\epsilon_i$. Numerical values of these parameters are obtained by fitting the spectrum. It noted that since we are treating $m_q$ as a free parameter, therefore, we must take of $\epsilon_1=1$.

\subsection{Potential model of hybrid mesons}\label{hybmodel}
In hybrid mesons, the gluonic field is in its excited state. In this case, quark antiquark potential can be written as following
\begin{equation}\label{Vhybrid}
V_{hybrid}(r)= V_{Q\bar{q}}(r)+V_g(r),
\end{equation}
where $V_{Q\bar{q}}(r)$ is the potential for conventional mesons appearing in Eq.~\eqref{con.potential}. $V_g(r)$ is the potential energy difference between the ground and the first excited state of gluonic field. Our proposed model of $V_g(r)$ is \cite{S.Noor},
\begin{equation}\label{Vg}
  V_g(r)= A e^{-Br^\nu} + \frac{\mu}{r}.
\end{equation}
Numerical values of parameters $A$, $B$, $\nu$ and $\mu$ are determined by fitting the model with recent lattice data taken from Fig. 2 of Ref. \cite{HybVdata}. The parameters are estimated by minimizing the $\chi^{2}$ function using a global numerical optimization algorithm. 
Our improved fitted values of $A$, $B$, $\nu$ and $\mu$ are 3.0552 GeV, 0.9369 GeV, 0.3881 and $-0.0025$ respectively. There is no physical motivation for using above model for fitting gluon potential except that it provides a nice fit to the lattice data.
\section{Mixed states}\label{mixing}
In $D$ and $D_s$ meson, charge conjugation is not a conserved quantum number, and states are described only by their $J^P$ values.
States having the same value of $J^P$ can mix. Except $0^-$ and $0^+$ all possible values of $J^P$ have mixed states. State mixing is categorized into two types, $S$-$D$ mixing and spin mixing.
  \subsection{$S$-$D$ mixing}
  It is mixing of spin triplet states $(^3L_{J}$$\leftrightarrow$$^3L'_{J})$ having different values of orbital angular momentum ($L$, $L'$) and same value of total angular momentum $J=L+1=L'-1$, so that $L'=L+2$.
  $^3S_1$ and $^3D_1$ states have the same values of $J^P$, therefore, they are mixed. Similarly $^3P_2$$\leftrightarrow$$^3F_2$, $^3D_3$$\leftrightarrow$$^3G_3$, $^3F_4$$\leftrightarrow$$^3H_4$, $^3G_5$$\leftrightarrow$$^3I_5$ and so on, all are referred as $S$-$D$ mixed states.
$S$-$D$ mixing occurs due to the tensor operator in the potential model, as this operator does not conserve orbital angular momentum.
  \subsection{Spin mixing}
  The spin-orbit interaction term in the potential model does not conserve total spin of quark and antiquark. It causes mixing between spin triplet and spin singlet states $(^1L_{L}$$\leftrightarrow$$^3L_{L})$, for which $J=L$. In this case, possible combinations of mixed states are $^1P_1$$\leftrightarrow$$^3P_1$, $^1D_2$$\leftrightarrow$$^3D_2$, $^1F_3$$\leftrightarrow$$^3F_3$, $^1G_4$$\leftrightarrow$$^3G_4$, $^1H_5$$\leftrightarrow$$^3H_5$ and so on.

\section{Relativistic wave equation}\label{rel wave eq}
In studying the  charmed and charmed strange mesons relativistic effects should be incorporated due to the presence of light quark. We use the relativistic wave equation to study the bound states of charmed and charmed strange mesons \cite{GI85}. The Schr$\mathrm{\ddot{o}}$dinger-type equation in relativistic dynamics is
\begin{equation}\label{rel.eq}
  H\psi(\mathbf{r})=E\psi(\mathbf{r}),
\end{equation}
where $\psi(\mathbf{r})$ and $E$ are wave function and energy of meson respectively. $H$ is effective Hamiltonian of meson:
 \begin{equation}
  H = \sqrt{\mathbf{p}^2+m_Q^2}+\sqrt{\mathbf{p}^2+m_{\bar{q}}^2}+V(\mathbf{r}),
\end{equation}
where $\mathbf{p}$ is center-of-mass momentum of quark such that $\mathbf{p}$ = $\mathbf{p}_Q$ = $-\mathbf{p}_{\bar{q}}$. $\mathbf{p}_Q$ and $\mathbf{p}_{\bar{q}}$ are momenta of quark and antiquark. $m_Q$ and $m_{\bar{q}}$ are mass of heavy quark (\emph{i.e.,} charm quark) and light antiquark respectively. $V(\mathbf{r})$ is the effective potential of bound state of quark antiquark system, discussed in Sec.~\ref{potential model}. We assumed our potential is central potential, $i.e.$, $V(\mathbf{r})\equiv V(r)$. For conventional mesons $V(r)$ $=$ $V_{Q\bar{q}}(r)$ (appearing in Eq.~\eqref{con.potential}) and for hybrid mesons $V(r)$ $=$ $V_{hybrid}(r)$ (appearing in Eq.~\eqref{Vhybrid}).

A conventional meson is a bound state of a quark and an antiquark. When the distance between them is large enough, the wave function vanishes. We assume the distance at which the wave function vanishes is $R$.
In order to solve Eq.~\eqref{rel.eq}, we find matrix representation of relativistic Hamiltonian $H$ using spherical Bessel function as basis. The diagonalization of the Hamiltonian matrix yields $\psi(\mathbf{r})$ and $E$.
For unmixed state (\emph{i.e.,} $0^{-}$, $0^{+}$) radial wave function is
\begin{equation}\label{RUmeq}
R_{l}(r) = \sum_{i=1}^{N} c_i\frac{a_{i}}{R} j_l\left(\frac{a_{i}r}{R}\right),
\end{equation}
where $c_i^{\;,}s$ are expansion coefficients, $j_l$ is the spherical Bessel function and $a_{i}$ is $i^{th}$ root of the spherical Bessel function such that $j_{l}(a_{i})$ = $0$.
In this basis the Eq.~\eqref{rel.eq} reduces to Ref.~\cite{ly2015}
\small
\begin{align}\label{lyEq}
\frac{2\Delta a_{i} a_{i}^{2}}{\pi R^{3}} & N_{li}^{2} \left(\sqrt{\left(\frac{a_{i}}{R}\right)^{2}+m_Q^{2}}+\sqrt{\left(\frac{a_{i}}{R}\right)^{2}+m_{\bar{q}}^{2}} \right)c_i+\sum_{j=1}^{N}\frac{a_{j}}{N_{li}^{2}a_{i}} \int_{0}^{R}drV(r)r^{2}j_l\left(\frac{a_ir}{R}\right)j_l\left(\frac{a_jr}{R}\right)c_j\notag\\
& =Ec_i,
\end{align}
\normalsize
where $N_{li}^{2}$ is radial integral of the spherical Bessel function give by
\begin{equation}
N_{li}^{2}=\int_{0}^{R}dr'r'^{2}j_l\left(\frac{a_ir'}{R}\right)^{2}.
\end{equation}
In Eq.~\eqref{lyEq}, the Hamiltonian matrix has the order $N\times N$. By diagonalising the Hamiltonian matrix, we get $N$ eigenvalues and eigenfunctions. Each eigenvalue and eigenfunction is associated with a particular value of the radial quantum number. For each eigen vector the resultant radial wave function is obtained from Eq.~\eqref{RUmeq}.
In case of mixed states, $\psi(\mathbf{r})$ is combination of two states.
For spin mixed states, $\psi(\mathbf{r})$ is
\begin{equation}\label{psiSpinMix}
\psi(\mathbf{r}) = \sum_{i=1}^{N} c_i^{(1)}\frac{a_i}{R}j_l\left(\frac{a_ir}{R}\right)\ket{^1L_L}_{\hat{r}} + \sum_{i=1}^{N} c_i^{(2)}\frac{a_i}{R}j_l\left(\frac{a_ir}{R}\right)\ket{^3L_L}_{\hat{r}},
\end{equation}
where $a_i$ is zeroth roots of the spherical Bessel function corresponding to $L$ orbital angular momentum. $c_i^{(1)}$ and $c_i^{(2)}$ are expansion coefficients associated to $\ket{^1L_L}_{\hat{r}}$ state and $\ket{^3L_L}_{\hat{r}}$ state respectively, which are defined as:
\begin{equation}\label{spintst1L}
  \ket{^1L_L}_{\hat{r}} = \ket{0,0}_s Y_{L,L}(\hat{r}),
\end{equation}
\begin{equation}\label{spintst3L}
  \ket{^3L_L}_{\hat{r}} = \sum_{m_s}C_{m_s}\ket{1,m_s}_s Y_{L,L-m_s}(\hat{r}).
\end{equation}
Where $C_{m_s}$ is Clebsch-Gorden coefficient.
In mixed states the Eq.~\eqref{rel.eq} is reduced to two coupled equations.
For spin mixed states the equations are:
 \small
 \begin{align}\label{spinly1}
  \frac{2}{\pi R^{3}} & \Delta a_i a_i^{2} N_{li}^{2} \left(\sqrt{\left(\frac{a_i}{R}\right)^{2}+m_Q^{2}}+\sqrt{\left(\frac{a_i}{R}\right)^{2}+m_{\bar{q}}^{2}} \right)c_i^{(1)}+\sum_{j=1}^{N}\frac{a_j}{N_{li}^{2}a_i} \int_{0}^{R}dr\left(\expect{V}_{11}c_j^{(1)}+\expect{V}_{12}c_j^{(2)}\right)\notag\\
  & r^{2}j_l\left(\frac{a_ir}{R}\right)j_l\left(\frac{a_jr}{R}\right)=Ec_i^{(1)},
  \end{align}
   \begin{align}\label{spinly2}
  \frac{2}{\pi R^{3}}& \Delta a_i a_i^{2} N_{li}^{2} \left(\sqrt{\left(\frac{a_i}{R}\right)^{2}+m_Q^{2}}+\sqrt{\left(\frac{a_i}{R}\right)^{2}+m_{\bar{q}}^{2}} \right)c_i^{(2)}+\sum_{j=1}^{N}\frac{a_j}{N_{li}^{2}a_i} \int_{0}^{R}dr\left(\expect{V}_{21}c_j^{(1)}+\expect{V}_{22}c_j^{(2)}\right)\notag\\
  &r^{2}j_l\left(\frac{a_ir}{R}\right)j_l\left(\frac{a_jr}{R}\right)=Ec_i^{(2)}.
  \end{align}
\normalsize
In this case the Hamiltonian matrix $H$ can be expressed through four $N\times N$ matrices $H_{11}$, $H_{12}$, $H_{21}$, and $H_{22}$ as following
\begin{equation}\label{hmixed}
  H =
\begin{pmatrix}
  H_{11} & H_{12} \\
  H_{21} & H_{22}
\end{pmatrix}.
\end{equation}

$H_{11}$ and $H_{12}$ are read off from Eq.~\eqref{spinly1}, and $H_{21}$ and $H_{22}$ from Eq.~\eqref{spinly2}.
 The kinetic energy of mesons contributes only to diagonal Hamiltonian matrices ($H_{11}, H_{22}$). $\expect{V}_{11}$, $\expect{V}_{12}$, $\expect{V}_{21}$ and $\expect{V}_{22}$ are potential energy matrix elements of $H_{11}$, $H_{12}$, $H_{21}$ and $H_{22}$ respectively. The definition of the potential energy matrix elements is mentioned in Table~\ref{thamMatrix}.
 It is noted that mixing of the states occurs due to off-diagonal Hamiltonian matrices ($H_{12}, H_{21}$). For spin mixed state, contribution to $\expect{V}_{12}$ and $\expect{V}_{21}$ come through spin-orbit terms of the potential energy.

In $S$-$D$ mixed states, $\psi(\mathbf{r})$ is
\begin{equation}\label{psiSDMix}
\psi(\mathbf{r}) = \sum_{i=1}^{N} c_i^{(1)}\frac{a_i}{R}j_l\left(\frac{a_ir}{R}\right)\ket{^3L_J}_{\hat{r}} + \sum_{i=1}^{N} c_i^{(2)}\frac{a'_i}{R}j_{l'}\left(\frac{a_i'r}{R}\right)\ket{^3L'_J}_{\hat{r}},
\end{equation}
where $a_i$ and $a_i'$ are zeroth roots of the spherical Bessel function corresponding to $L$ and $L'$ orbital angular momentum respectively.
In $S$-$D$ mixed states, $c_i^{(1)}$ and $c_i^{(2)}$ are expansion coefficients associated to $\ket{^3L_J}_{\hat{r}}$ state and $\ket{^3L'_J}_{\hat{r}}$ state respectively, which are defined as:
\begin{equation}\label{sd3L}
  \ket{^3L_J}_{\hat{r}} = \ket{1,1}_s Y_{L,L}(\hat{r}),
\end{equation}
\begin{equation}\label{spintst3L}
  \ket{^3L'_J}_{\hat{r}} = \sum_{m_s}C_{m_s}\ket{1,m_s}_s Y_{L+2,L+1-m_s}(\hat{r}).
\end{equation}

For $S$-$D$ mixed states, we have following set of equations:
\footnotesize
%\small
 \begin{align}\label{sdly1}
  \frac{2}{\pi R^{3}} & \Delta a_i a_i^{2} N_{li}^{2}  \left(\sqrt{\left(\frac{a_i}{R}\right)^{2}+m_Q^{2}}+\sqrt{\left(\frac{a_i}{R}\right)^{2}+m_{\bar{q}}^{2}} \right)c_i^{(1)}  +\sum_{j=1}^{N}\frac{a_j}{N_{li}^{2}a_i} \int_{0}^{R}dr\expect{V}_{11}r^{2}j_l\left(\frac{a_ir}{R}\right)j_l\left(\frac{a_jr}{R}\right)c_j^{(1)}\notag\\
  & +\sum_{j=1}^{N}\frac{a_j'}{N_{li}^{2}a_i} \int_{0}^{R}dr\expect{V}_{12}r^{2}j_l\left(\frac{a_ir}{R}\right)j_{l'}\left(\frac{a_j'r}{R}\right)c_j^{(2)}=Ec_i^{(1)},
  \end{align}
  \begin{align}\label{sdly2}
 \frac{2}{\pi R^{3}} & \Delta a_i' a_i'^{2} N_{l'i}^{2} \left(\sqrt{\left(\frac{a_i'}{R}\right)^{2}+m_Q^{2}}+\sqrt{\left(\frac{a_i'}{R}\right)^{2}+m_{\bar{q}}^{2}} \right)c_i^{(2)}+\sum_{j=1}^{N}\frac{a_j}{N_{l'i}^{2}a_i'} \int_{0}^{R}dr\expect{V}_{21}r^{2}j_{l'}\left(\frac{a_i'r}{R}\right)j_l\left(\frac{a_jr}{R}\right)c_j^{(1)}\notag\\
  & +\sum_{j=1}^{N}\frac{a_j'}{N_{l'i}^{2}a_i'} \int_{0}^{R}dr\expect{V}_{22}r^{2}j_{l'}\left(\frac{a_i'r}{R}\right)j_{l'}\left(\frac{a_j'r}{R}\right)c_j^{(2)}=Ec_i^{(2)}.
 \end{align}
 \normalsize
 Once again these equations can be used to define Hamiltonian matrix $H$ in terms of four $N\times N$ matrices $H_{11}$, $H_{12}$, $H_{21}$ and $H_{22}$.
 It is noted that for $S$-$D$ mixed states, contribution to $\expect{V}_{12}$ and $\expect{V}_{21}$ comes through tensor operator of the potential.
 \begin{table}[t]
\caption{The definition of potential energy matrix elements for spin and $S$-$D$ mixed states.}
\begin{tabular}{c  c c}
\hline\hline
\;Potential energy matrix elements& \hspace{.05 in} Spin mixed state  & \hspace{.05 in} $S$-$D$ mixed state  \\ [1ex]
 \hline
 \hline
  $\expect{V}_{11}$ &  $\braopket{^1L_L}{V}{^1L_L}$ & $\braopket{^3L_J}{V}{^3L_J}$\\
  $\expect{V}_{12}$ &  $\braopket{^1L_L}{V_{so}}{^3L_L}$ &  $\braopket{^3L_J}{V_{T}}{^3L'_J}$\\
  $\expect{V}_{21}$ &  $\braopket{^3L_L}{V_{so}}{^1L_L}$ & $\braopket{^3L'_J}{V_{T}}{^3L_J}$\\
  $\expect{V}_{22}$ &  $\braopket{^3L_L}{V}{^3L_L}$ & $\braopket{^3L'_J}{V}{^3L'_J}$\\
 \hline\hline
\end{tabular}
  \label{thamMatrix}
\end{table}

\begin{figure}[ht]
  \centering
  \includegraphics[scale=0.9]{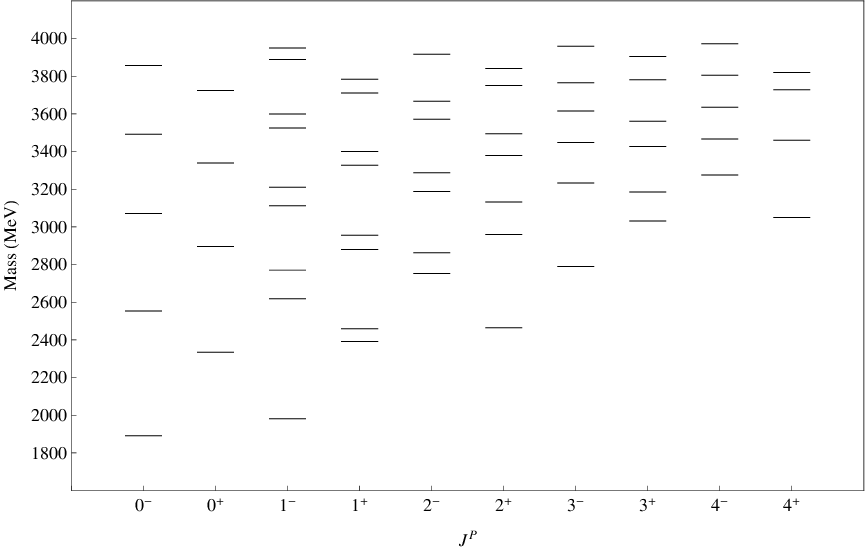}
  \caption{Conventional charmed mesons spectrum.}\label{fD}
  \vspace*{\floatsep}% https://tex.stackexchange.com/q/26521/5764
\includegraphics[scale=0.9]{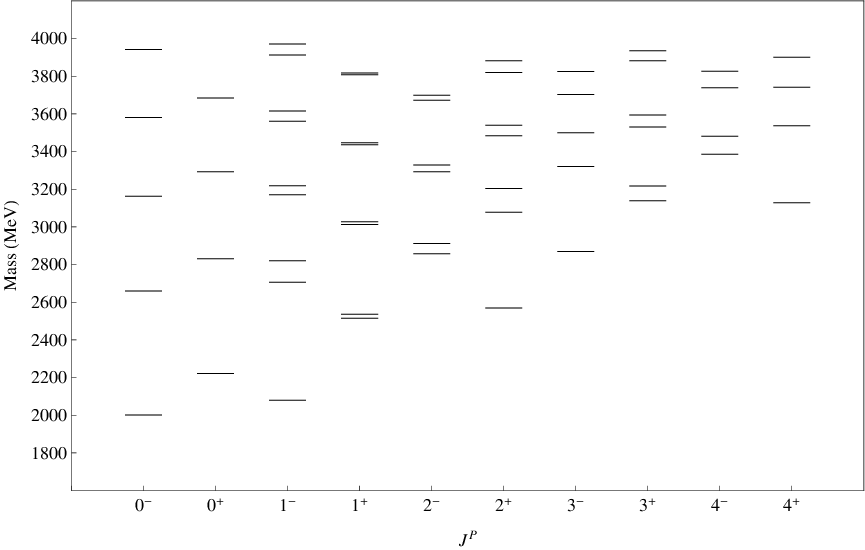}
  \caption{Conventional charmed strange mesons spectrum.}\label{fDs}
\end{figure}

\begin{figure}[ht]
  \centering
  \includegraphics[scale=0.9]{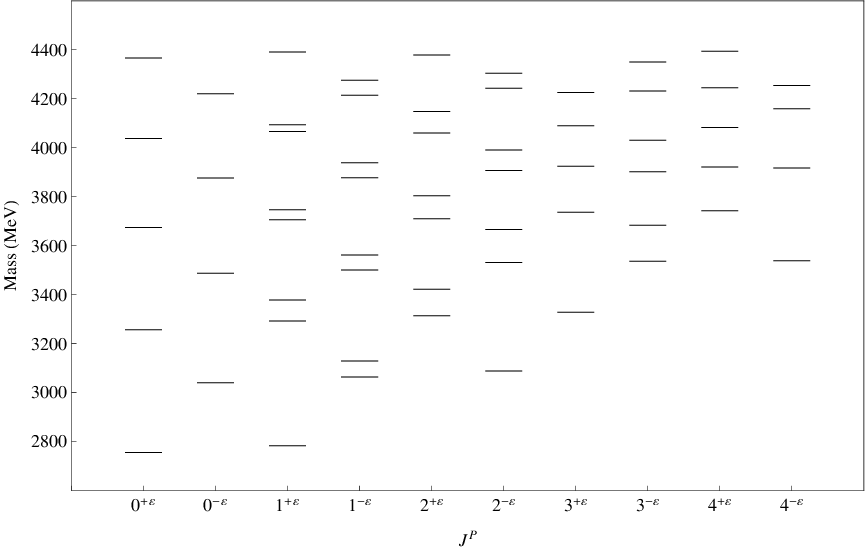}
  \caption{Hybrid charmed mesons spectrum.}\label{fhD}
  \vspace*{\floatsep}% https://tex.stackexchange.com/q/26521/5764
\includegraphics[scale=0.9]{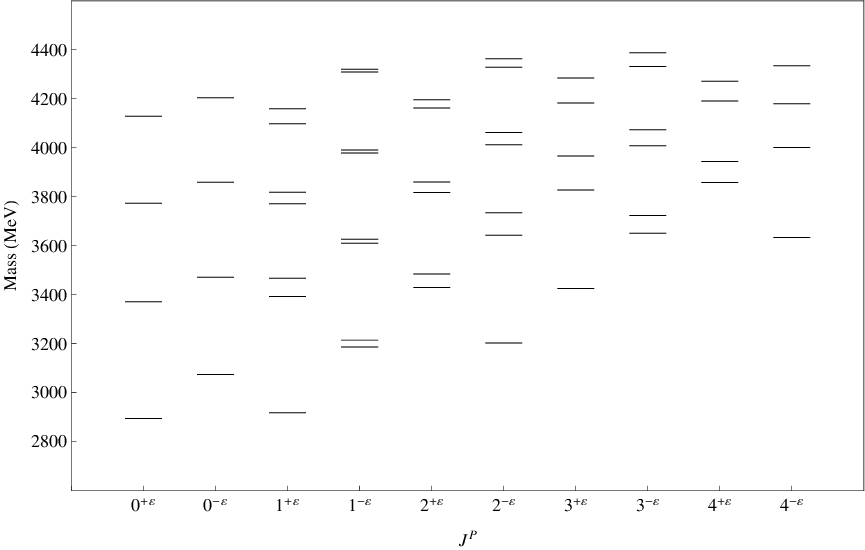}
  \caption{Hybrid charmed strange mesons spectrum.}\label{fhDs}
\end{figure}

Our numerical results also depend upon the order of the matrix $N$. We observe that for a large value of $N$, dependency is negligible. In our calculations we take $N$ = 50 and $R=25$ $\mathrm{GeV^{-1}}$.
%To incorporate relativistic and non-perturbative effects, mass of the light quark in spin dependent part of the potential model is calculated in the same way as did in Refs~\cite{ly,ly2015}. Such that $\tilde{m}_{qi}$ = $\tilde{m}_{\bar{q}}\epsilon_i$, $i=1, 2, 3, 4$.
Numerical values of quark masses and other parameters of the conventional potential model are extracted from fitting the spectrum to experimentally known states. We used $14$ experimentally well-established states for fitting. The states that we used for fitting parameter values are mentioned in Table \ref{TspectrumCmp}.
For fitting the spectrum, we used the differential evolution method \cite{diffEvo}. Differential evolution (DE) is a stochastic global optimization algorithm. In consecutive independent trials, DE successfully converged to the global minimum of the function. By minimizing the $\chi^{2}$ function using the DE algorithm, we find out the best fit values of parameters.
     Our fitted values of parameters are:
$m_{c}$ = 1.47 $\pm$ 0.01 GeV, $m_{u/d}$ = 0.020 $\pm$ 0.001 GeV, $\tilde{m}_{u/d}$ = 0.430 $\pm$ 0.001 GeV, $m_{s}$ = 0.336 $\pm$ 0.007 GeV, $\tilde{m}_{s}$ = 0.5967 $\pm$  0.00046 GeV, $\alpha_s^{c}$ = 0.240 $\pm$ 0.025, $b$ = 0.211 $\pm$ 0.002 $\mathrm{GeV^2}$, $c$ = -0.4321 $\pm$ 0.0041 GeV, $\sigma_0$ = 0.5221 $\pm$ 0.013 GeV, $s_0$ = 0.91 $\pm$ 0.15, $\epsilon_2$ = 1.5 $\pm$ 0.59, $\epsilon_3$ = 1.5 $\pm$ 0.1, $\epsilon_4$ = 1.31 $\pm$ 0.14. 
Our predicted spectra of conventional charmed and charmed strange mesons is given in Fig.~\ref{fD} and Fig.~\ref{fDs}. In Fig.~\ref{fhD} and Fig.~\ref{fhDs}, we present our calculated spectrum of hybrid charmed mesons and hybrid charmed strange mesons.  Comparison of our calculated spectrum of conventional $D$ and $D_s$ mesons with other works is given in Tables \ref{TspectrumCmp} and \ref{TspectrumCmp1}.
%figure conventional D Ds
%fig hybrid D Ds
%table comparison of our calculated mass results in MeV with other works
\begin{table}[b]
\caption{Comparison of our calculated spectrum (in MeV) of conventional $D$ and $D_s$ mesons with other works.}
\begin{tabular}{ccccccc}
 \hline\hline
\;\;\;Meson\;\;\;&\;\;\;$J^P$\;\;\; & \;\;\;Our Mass Results\;\;\;\;& GI Model\cite{godfrey2016}\;\;\;\;& \;\;\;\;\cite{ly2015}\;\;\;\;&\;\;\;\;\cite{chen2020}&\;\;\;\;PDG\cite{PDG}\\
\hline \hline
Charmed strange& mesons:\\
$D_s^\pm$           & $0^-$ & 2001.36 & 1979 & 1963 & 1964 & 1968.35 $\pm$ 0.07 \\
$D_s^{*\pm}$        & $1^-$ & 2079.63 & 2129 & 2103 & 2102 & 2112.2 $\pm$ 0.4 \\
$D_{s1}(2536)^\pm$  & $1^+$ & 2535.92 & 2556 & 2533 & 2510 & 2535.11 $\pm$ 0.06 \\
$D_{s2}^*(2573)$    & $2^+$ & 2569.18 & 2592 & 2573 & 2548 & 2569.1 $\pm$ 0.8 \\
$D_{s1}^*(2700)^\pm$& $1^-$ & 2706.22 & 2732 & 2709 & 2680 & 2714 $\pm$ 5 \\
$D_{s1}^*(2860)^\pm$& $1^-$ & 2820.59 & 2899 & 2803 & 2771 & 2859 $\pm$ 27 \cite{Ds1(2860), Ds1(2860)(2)}\\
$D_{s3}^*(2860)^\pm$  & $3^-$ & 2869.84 & 2917 & 2853 & 2816 & 2860 $\pm$ 7  \\
Charmed mesons:\\
$D^0$          & $0^-$ & 1890.66 & 1877 & 1867 & &1864.83 $\pm$ 0.05   \\
$D^*(2007)^0$  & $1^-$ & 1980.63 & 2041 & 2004 & &2006.85 $\pm$ 0.05   \\
$D_0^*(2300)^{0}$  & $0^+$ & 2334.64 & 2399 & 2302 & &2343 $\pm$ 10 \\
$D_1(2430)^{0}$    & $1^+$ & 2458.56 & 2456 & 2415 & & 2412 $\pm$ 9 \\
$D_1(2420)^0$  & $1^+$ & 2391.87 &  2467 & 2429 & & 2422.1 $\pm$ 0.6 \\
$D_2^*(2460)^{0}$  & $2^+$ & 2464.39  & 2502 & 2468 & & $2461.1_{-0.8}^{+0.7}$ \\
$D_3^*(2750)$  & $3^-$ & 2789.59 & 2833 & 2754 & &2763.1 $\pm$ 3.2  \\
  \hline
  \hline
  \end{tabular}
  \label{TspectrumCmp}
\end{table}
%-----------------------------------------------------------------
%table comparison of our calculated mass results in MeV with other works
\begin{table}[t]
\caption{Comparison of our calculated spectrum (in MeV) of conventional $D$ and $D_s$ mesons with others works and experimental data.}
\begin{tabular}{ccccccc}
 \hline\hline
Meson\;\;\;&\;\;\;$J^P$\;\;\; & \;\;\;Our Results\;\;\;\;& GI Model\cite{godfrey2016}\;\;\;\;& \cite{ly2015}\;\;\;\;&\;\;\;\;\cite{chen2020}&\;\;\;\;Exp. Data\\
\hline \hline
Charmed strange& mesons:\\
$D_{s0}^*(2590)^+$    & $0^-$ & 2659 & 2673 & 2582 & 2557 & 2591$\pm$9 \cite{Ds(2590)} \\
$D_{sJ}^*(3040)^\pm$ &  $1^+$, $2^+$ & 3026.94, 3077.36 & & & & $3044_{-9}^{+31}$ \cite{Ds1(2860)(2)}\\
Charmed mesons:\\
$D_0(2550)^0$  & $0^-$ & 2553.34 & 2581 & 2483 & & 2549 $\pm$ 19 \cite{D(2550)1, D(2550)} \\
$D_1^*(2600)^0$  & $1^-$ & 2618.66 & 2643 & 2613 & & 2627 $\pm$ 10 \cite{D(2550), D1(2600)1, D1(2600)}\\
$D_2(2740)^0$  & $2^-$ & 2752.03 & 2816 & 2737 & & 2747 $\pm$ 6 \cite{D1(2600), D2(2740)}\\
$D_1^*(2760)^0$  & $1^-$ & 2770.32 & 2817 & 2785 & & 2781 $\pm$ 22 \cite{D1(2760)}\\
$D(3000)^0$  & $3^+$, $3^-$, $4^-$\;\; & 3184.76, 3233.29, 3275.27 &  &  & & 3214 $\pm$ 60 \cite{D1(2600)1, D2(2740)} \\
  \hline
  \hline
  \end{tabular}
  \label{TspectrumCmp1}
\end{table}
\section{Radiative transitions}\label{radiative transitions}
Radiative transitions play a vital role in the identification of meson states.
$E1$ radiative partial width between initial and final mesons in non-relativistic quark model is \cite{ishrat19}
%\cite{kwong1988}
\begin{equation}\label{e1}
\Gamma(n^{2S+1}L_J \rightarrow n'^{2S'+1}L'_{J'} + \gamma) = \frac{4}{3}\expect{e_Q}^2\alpha\omega^3C_{fi}\delta_{SS'}\delta_{L,L'\pm1}|\braopket{n'^{2S'+1}L'_{J'}}{r}{n^{2S+1}L_J}|^2\frac{E_f}{M_i},
\end{equation}
where $\alpha =$ \large$\frac{1}{137}\;$\normalsize is electromagnetic fine structure constant, $\omega$ is photon energy, $\ket{n^{2S+1}L_J}$ is initial meson state, $\ket{n'^{2S'+1}L'_{J'}}$ is final meson state, $E_f$ is the energy of final meson state and $M_i$ is mass of initial meson. $C_{fi}$ is angular momentum matrix element defined as
\begin{equation*}
C_{fi} = max(L,L')(2J' + 1) \begin{Bmatrix}L'&J'&S\\J&L&1 \end{Bmatrix}^2,
\end{equation*}
where \tiny$\begin{Bmatrix}.&.&.\\.&.&. \end{Bmatrix}\;$\normalsize is a $6-j$ symbol. In Eq.~\eqref{e1}, $\expect{e_Q}$ is
\begin{equation}
\expect{e_Q} = \frac{m_ce_q - m_{q}e_{\overline{c}}}{m_{q} + m_c}.
\end{equation}
Here $m_c$ and $m_{q}$ are constituent mass of charm quark and  constituent mass  of light ($q$) quark respectively. For $D$ meson, $q$ = $u$/$d$ whereas for $D_s$ meson $q$ = $s$. $e_{\overline{c}}$ = $-$2/3 is electric charge of anti-charm quark in units of $|e|$. $e_u$ = +2/3, $e_d$ = $-$1/3 and $e_s$ = $-$1/3 are electric charge of up, down and strange quark respectively in units of $|e|$.

Matrix elements $\braopket{n'^{2S'+1}L'_{J'}}{r}{n^{2S+1}L_J}$ are calculated using the relativistic quark model wave functions obtained in Sec. \ref{rel wave eq}. In E1 transitions, relativistic corrections are implicitly included due to the contribution of spin dependent interactions in the Hamiltonian.
\\
$M1$ radiative partial width for $S$-wave meson bound state is \cite{godfrey2016}
\begin{align}\label{m1}
\Gamma & (n^{2S+1}L_J \rightarrow n'^{2S'+1}L_{J'} + \gamma)\notag\\
& = \frac{\alpha}{3}\omega^3(2J^{'}+1)\delta_{S,S'\pm1}|\braopket{f}{\frac{e_q}{m_q}j_0\left(\frac{m_c}{m_q+m_c}\omega r\right)-\frac{e_{\bar{c}}}{m_c}j_0\left(\frac{m_q}{m_q+m_c}\omega r\right)}{i}|^2,
\end{align}
%\normalsize
where $j_0$ is the spherical Bessel function.
%Due to the contribution of spin dependent part of the potential model, relativistic dynamics is incorporated in radiative transitions.
Electromagnetic transitions are sensitive to the internal structure of mesons, particularly in mixed states. A mixed state is a combination of two states, only that state(s) contributes to $E1$ and $M1$ transitions, which fulfill the selection rules of these transitions.
For calculating decay width of $M1$ transitions of conventional S-wave meson bound state, selection rules of $M1$ transitions and Eq.~\eqref{psiSDMix} implies that in this case only $1^{-}\leftrightarrow 0^{-}$ transition allowed further only $^{3}S_{1}$ component of $1^{-}$ state contributes to decay width. To calculate decay width of $E1$ transitions of conventional meson bound states ($L<3$), selection rules of $E1$ transitions and Eqs.~(\ref{psiSpinMix}, \ref{psiSDMix}) indicate that only four transitions contribute, i.e., $1^{+}\leftrightarrow 0^{-}$, $1^{-}\leftrightarrow 0^{+}$, $1^{-}\leftrightarrow 1^{+}$ and $2^{-}\leftrightarrow 1^{+}$.
In $1^{+}\leftrightarrow 0^{-}$ transition, only $^{1}P_1$ component of $1^{+}$ state fulfil selection rule of $E1$ transitions. In $1^{-}\leftrightarrow 0^{+}$ transition both components of $1^{-}$ state participate. In $1^{-}\leftrightarrow 1^{+}$, both components of the initial and final states take part except the $^{1}P_{1}$ component of $1^{+}$ state. Finally, in $2^{-}\leftrightarrow 1^{+}$ transition both components of the initial and final states contribute.

\section{Results and conclusions}\label{results}
In this work we present a rich spectrum of conventional and hybrid charmed and charmed strange mesons. States having the same value of $J^P$ are mixed. We have incorporated both spin and $S$-$D$ mixing through the off-diagonal matrix elements of the Hamiltonian. In Fig.~\ref{fD} to \ref{fhDs}, state mixing phenomena are explicitly visualized by the spectrum pattern. Ref.~\cite{godfrey2016} include spin mixing effects although they ignore $S$-$D$ mixing in their calculations. In Ref.~\cite{ly2015}, spin dependent Hamiltonian part is added perturbatively after solving the eigenequation.
In this work, we use the complete Hamiltonian (including spin dependent part) in the eigenequation. This allows us to study both spin and $S$-$D$ mixing effects without resorting to perturbation theory.
%two charmed strange mesons-----------
%Theoretical predictions of masses of two charmed strange mesons (\emph{i.e.,} $D^*_{s0}(2317)^{\pm}$, $D_{s1}(2460)^{\pm}$) are larger than experimentally observed masses. A lot of work is done in literature to identify the nature of these mesons. In some literature~\cite{tetraquark1,tetraquark2} these states are interpreted as tetraquarks, while in some places $D^*_{s0}(2317)^{\pm}$ and $D_{s1}(2460)^{\pm}$ are taken as $DK$ and $DK^{*}$ molecule states respectively~\cite{dkbarn2003,dkchen2004}. In Refs.~\cite{godfrey2016, ly, ly2015, chen2020}, these states have been studied as conventional mesons. In this work, we identify these states as conventional mesons. We observed that by incorporating spin and $S$-$D$ mixing, the difference between theoretical and experimental masses of these states is significantly reduced. In our calculations, the difference is much lower than Ref.~\cite{godfrey2016}.
%--------------------

Our calculated spectrum of conventional $D$ and $D_s$ mesons is reported in Table \ref{tConvSpectrum} with increasing order of value of mass.
$n$ is number of eigenvalues associated with specific value of $J^P$. For unmixed states, it is visible that $0^+$ ($^3P_0$) states have larger mass as compared to $0^-$ ($^1S_0$) states due to spin and orbital excitation. In the case of $S$-$D$ mixed states, higher orbital excited states (\emph{e.g.,} $2^+$, $^3P_2$$\leftrightarrow$$^3F_2$) have large mass compared to lower one ($e.g.,$ $1^-$, $^3S_1$$\leftrightarrow$$^3D_1$). Same is observed for spin singlet and spin triplet mixed states. States with higher orbital excitations ($e.g.,$ $2^-$, $^1D_2$$\leftrightarrow$$^3D_2$) have large mass
in comparison to lower orbital excited states ($e.g.,$ $1^+$, $^1P_1$$\leftrightarrow$$^3P_1$). We also compare our calculated spectrum of conventional $D$ and $D_s$ mesons with experimental results and other works in Tables \ref{TspectrumCmp} and \ref{TspectrumCmp1}. Our results have good agreement with other works as well as with experimental data.
We used meson states in Table \ref{TspectrumCmp} for fitting the parameters value. Table \ref{TspectrumCmp1} shows the efficiency of our results. In Table \ref{TspectrumCmp1}, we identify $D_{sJ}^*(3040)^\pm$ and $D(3000)^0$  as $1^+$ and $3^-$ state. We tentatively identify $D_{sJ}^*(3040)^\pm$ and $D(3000)^0$ as $2^+$ and $3^+$ state. Possibility of $D(3000)^0$ could be $4^-$ state. Since the nature of $D^*_{s0}(2317)^{\pm}$ and $D_{s1}(2460)^{\pm}$ states are not clear as discussed in Sec. \ref{intro}, we do not include these states in the fitting procedure. Moreover, we do not find any good match to these states in conventional as well as hybrid mass spectrum. This result is consistent with hitherto exotic nature of $D_{s0}^*(2317)$ and $D_{s1}(2460)$ mesons.

In Table \ref{tHybridSpectrum}, we report our computed masses of hybrid $D$ and $D_s$ mesons.
For hybrid candidates, parity condition is different to that for conventional mesons; this is due to the excited gluonic field contribution as discussed in Sec.~\ref{intro}.
It is observed that hybrid states have greater masses comparative to that of conventional ones. This happen due to gluonic excitations in $\Pi_u$ states.  In Table \ref{tConvE1} and Table \ref{tHybridE1} we have reported our calculated $E1$ transitions of conventional to conventional and hybrid to hybrid mesons respectively. We also present  $M1$ transitions of conventional to conventional and hybrid to hybrid mesons in Table \ref{tConvM1} and Table \ref{tHybridM1} respectively. We observed that the decay width of $D^0(c\bar{u})$ states is large as compared to $D^\pm(c\bar{d})$ states except few states in M1 transitions.
In Table \ref{tcompRadiative}, we compare our calculated radiative decay width of conventional mesons with other works and experimental results. Our results are consistent with other works except for a few states of $D^0$ mesons. We expect our predicted results will be useful in detection of new charmed and charmed strange mesons.

%table conventional D Ds
%table conventional D Ds
\LTcapwidth=5in
\begin{longtable}{cccc}
\caption{Our calculated conventional $D$ and $D_s$ meson masses (in MeV) for $J^P$ = $0^-$, $0^+$, $1^-$, $1^+$, $2^-$, $2^+$, $3^-$, $3^+$, $4^-$ and $4^+$.}\\
\hline\hline
\label{tConvSpectrum}
%First entry & \textbf{Second entry} & \textbf{Third entry} & \textbf{Fourth entry} \\
\;\;\;$J^P$\;\;\; & \;\;\;\;\;$n$\;\;\;\;\;\;\;& \;$D$ Meson\;\;\;\;&\;\;\;  $D_s$ Meson\\
\hline\hline
\endfirsthead
\multicolumn{4}{c}%
{\tablename\ \thetable\ -- \textit{Continued from the previous page}} \\
\hline\hline
%\textbf{First entry} & \textbf{Second entry} & \textbf{Third entry} & \textbf{Fourth entry} \\
\;\;\;$J^P$\;\;\; & \;\;\;\;\;$n$\;\;\;\;\;\;& \;$D$ Meson\;\;\; & \;\;\; $D_s$ Meson\\
\hline\hline
\endhead
\hline \multicolumn{4}{r}{\textit{Continued on the next page}} \\
\endfoot
%\hline
\endlastfoot
$0^-$ & 1 & 1890.66   & 2001.36  \\
      & 2 &  2553.34   & 2659.00 \\
      & 3 &  3070.94  & 3162.90 \\
      & 4 & 3491.35    & 3581.01 \\
 \hline
$0^+$ & 1 &  2334.64   & 2221.26 \\
      & 2 &  2895.60   & 2830.25  \\
      & 3 &  3339.03   & 3292.43 \\
      & 4 &  3723.79   &  3684.70\\
 \hline
$1^-$ & 1 &  1980.63  & 2079.63 \\
      & 2 &  2618.66   & 2706.22 \\
      & 3 &  2770.32   & 2820.59  \\
      & 4 &   3111.91  & 3170.35 \\
      & 5 &  3210.21   & 3218.06 \\
      & 6 &  3524.85   & 3561.34 \\
      & 7 &  3599.51   & 3614.62 \\
      & 8 &  3888.75  &  3912.91 \\
 \hline
$1^+$ & 1 & 2391.87   & 2514.83  \\
      & 2 & 2458.56   & 2535.92\\
      & 3 & 2880.36    & 3012.05 \\
      & 4 & 2955.80   & 3026.94 \\
      & 5 & 3327.55   & 3435.81 \\
      & 6 &   3400.51 & 3446.24  \\
      & 7 & 3710.89   & 3807.50 \\
      & 8 & 3784.40    & 3816.56 \\
 \hline
$2^-$ & 1 & 2752.03    & 2857.35 \\
      & 2 & 2862.13    & 2911.54 \\
      & 3 & 3188.32    & 3292.05  \\
      & 4 & 3287.53   & 3327.83 \\
      & 5 & 3571.28    & 3672.60 \\
      & 6 & 3666.60   & 3699.38 \\
      & 7 & 3916.99    & 4014.94 \\
      & 8 & 4008.79    & 4037.37 \\
 \hline
 $2^+$ & 1 &  2464.39   & 2569.18 \\
       & 2 &  2959.26   & 3077.36 \\
       & 3 &  3131.47   & 3203.58 \\
       & 4 &   3379.56  & 3483.48 \\
       & 5 &  3494.62  & 3539.61 \\
       & 6 &  3750.46  & 3819.04 \\
       & 7 &  3840.78  & 3881.70 \\
       & 8 &  4086.20  & 4130.11 \\
 \hline
 $3^-$ & 1 & 2789.59  & 2869.84 \\
      & 2 & 3233.29   & 3320.75 \\
      & 3 & 3448.43   & 3499.65  \\
      & 4 & 3615.56   & 3702.96 \\
      & 5 & 3765.23    & 3825.09 \\
      & 6 & 3959.01   & 4053.21 \\
      & 7 &  4075.86   & 4114.68 \\
      & 8 &  4274.05  & 4363.43 \\
 \hline
 $3^+$ & 1 & 3031.60  & 3138.04 \\
      & 2 &  3184.76  & 3216.67 \\
      & 3 &  3427.33   & 3529.81 \\
      & 4 & 3560.11   & 3594.36 \\
      & 5 & 3781.48   & 3881.76 \\
      & 6 &  3904.52  & 3935.16  \\
      & 7 &  4105.19  & 4203.53 \\
      & 8 &  4222.14  & 4248.79 \\
 \hline
 $4^-$ & 1 & 3275.27   & 3385.16 \\
      & 2 &  3466.85  & 3480.50 \\
      & 3 & 3634.65   & 3738.40 \\
      & 4 & 3804.95   & 3826.51 \\
      & 5 & 3971.98   & 4073.56 \\
      & 6 & 4125.85   & 4149.44 \\
      & 7 & 4278.55   & 4376.37  \\
      & 8 & 4424.86   & 4445.97 \\
 \hline
 $4^+$ & 1 & 3050.12  & 3128.49  \\
      & 2 &  3460.14  & 3536.36 \\
      & 3 &  3727.59  & 3740.77 \\
      & 4 &  3819.37   & 3900.19 \\
      & 5 &  4021.00  & 4050.96 \\
      & 6 &  4144.89  & 4230.53 \\
      & 7 &  4303.97  & 4357.16 \\
      & 8 &  4445.31  & 4534.66 \\
 \hline\hline
 %\label{tConvSpectrum}
\end{longtable}

%==========================================================================================
%table Hybrid Spectrum D Ds

\LTcapwidth=5in
\begin{longtable}{cccc}
\caption{Our calculated hybrid $D$ and $D_s$ meson masses (in MeV) for $J^P$ =  $0^{-\varepsilon}$, $0^{+\varepsilon }$, $1^{-\varepsilon}$, $1^{+\varepsilon}$, $2^{-\varepsilon}$, $2^{+\varepsilon}$, $3^{-\varepsilon}$, $3^{+\varepsilon}$, $4^{-\varepsilon}$ and $4^{+\varepsilon}$. }\\
\hline\hline
\label{tHybridSpectrum}
%First entry & \textbf{Second entry} & \textbf{Third entry} & \textbf{Fourth entry} \\
\;\;\;$J^P$\;\;\; & \;\;\;\;\;$n$\;\;\;\;\;\;\;& \;Hybrid $D$ Meson\;\;\;\;&\;\;\;  Hybrid $D_s$ Meson\\
\hline\hline
\endfirsthead
\multicolumn{4}{c}%
{\tablename\ \thetable\ -- \textit{Continued from the previous page}} \\
\hline\hline
%\textbf{First entry} & \textbf{Second entry} & \textbf{Third entry} & \textbf{Fourth entry} \\
\;\;\;$J^P$\;\;\; & \;\;\;\;\;$n$\;\;\;\;\;\;& \;Hybrid $D$ Meson\;\;\; & \;\;\; Hybrid $D_s$ Meson\\
\hline\hline
\endhead
\hline \multicolumn{4}{r}{\textit{Continued on the next page}} \\
\endfoot
%\hline
\endlastfoot
$0^{+\varepsilon}$ & 1 & 2755.01   & 2894.34  \\
      & 2 & 3256.86   & 3371.39 \\
      & 3 &  3674.27  & 3773.52 \\
      & 4 & 4038.16   & 4128.21 \\
 \hline
$0^{-\varepsilon}$ & 1 & 3040.45   & 3073.69 \\
      & 2 & 3487.44   & 3471.19 \\
      & 3 & 3876.36   & 3858.93 \\
      & 4 & 4220.98   & 4204.41  \\
 \hline
$1^{+\varepsilon}$ & 1 & 2783.31   & 2917.32 \\
      & 2 & 3291.98   & 3392.35 \\
      & 3 &  3377.88  & 3467.12 \\
      & 4 &  3705.86  & 3771.09 \\
      & 5 &  3747.28  & 3818.40  \\
      & 6 &  4066.32  & 4098.38 \\
      & 7 & 4094.33   & 4159.38 \\
      & 8 & 4391.27   & 4406.70 \\
 \hline
$1^{-\varepsilon}$ & 1 & 3063.57   & 3186.10 \\
      & 2 & 3129.03   & 3214.35 \\
      & 3 &  3501.03   & 3610.14 \\
      & 4 & 3561.83   & 3626.80 \\
      & 5 &  3877.47  & 3978.27 \\
      & 6 & 3938.51   & 3991.02 \\
      & 7 & 4214.71   & 4309.78 \\
      & 8 &  4275.42  & 4320.49 \\
 \hline
$2^{+\varepsilon}$ & 1 & 3313.71   & 3429.11 \\
      & 2 & 3422.03   & 3484.54  \\
      & 3 & 3710.06   & 3817.38 \\
      & 4 & 3803.52   & 3860.47 \\
      & 5 & 4060.25   & 4162.55 \\
      & 6 & 4148.00   & 4196.43 \\
      & 7 & 4378.76   & 4476.95 \\
      & 8 & 4462.95   & 4504.55 \\
 \hline
 $2^{-\varepsilon}$ & 1 & 3088.68   & 3202.11  \\
       & 2 &  3531.70  & 3642.38 \\
       & 3 & 3666.11   & 3734.10  \\
       & 4 & 3907.27   & 4012.51 \\
       & 5 &  3991.32  & 4062.51 \\
       & 6 &  4243.29   & 4329.07 \\
       & 7 &  4304.42  & 4363.03 \\
       & 8 &  4550.91  & 4612.00 \\
 \hline
 $3^{+\varepsilon}$ & 1 & 3328.26  & 3425.32 \\
      & 2 & 3736.37   & 3827.75 \\
      & 3 &  3924.76  & 3966.53 \\
      & 4 & 4089.99    & 4182.79 \\
      & 5 & 4225.45   & 4284.57 \\
      & 6 & 4409.37   & 4503.65 \\
      & 7 & 4512.24    & 4572.02 \\
      & 8 &  4703.77  & 4797.78 \\
 \hline
 $3^{-\varepsilon}$ & 1 & 3536.73  & 3650.73 \\
      & 2 & 3683.47   & 3723.53 \\
      & 3 & 3902.08   & 4008.45 \\
      & 4 & 4030.85   & 4073.06 \\
      & 5 & 4231.61   & 4331.70 \\
      & 6 & 4349.85   & 4387.91 \\
      & 7 & 4534.57   & 4619.30 \\
      & 8 & 4645.99   & 4671.07 \\
 \hline
 $4^{+\varepsilon}$ & 1 & 3742.56  & 3858.08 \\
      & 2 & 3921.79   & 3943.51 \\
      & 3 & 4082.66   & 4191.13 \\
      & 4 & 4244.97   & 4271.20 \\
      & 5 & 4394.41   & 4497.79 \\
      & 6 & 4544.24   & 4572.21 \\
      & 7 & 4683.88   & 4783.46 \\
      & 8 & 4824.56   & 4852.37 \\
 \hline
 $4^{-\varepsilon}$ & 1 & 3538.36  & 3633.01 \\
      & 2 & 3917.60   & 4000.87 \\
      & 3 & 4159.43   & 4179.46 \\
      & 4 & 4254.05   & 4335.09 \\
      & 5 &  4446.62  & 4483.56 \\
      & 6 & 4560.83   & 4641.67 \\
      & 7 & 4715.93   & 4763.97 \\
      & 8 & 4845.22    & 4927.05 \\
 \hline\hline
 %\label{tConvSpectrum}
\end{longtable}

%=================================================================================================
%table Conventional e1 radiative transition
\LTcapwidth=\textwidth
\begin{longtable}{ccccc}
\caption{Partial widths for $E1$ decays (in KeV) of conventional $D$ and $D_s$ mesons.}\\
\hline\hline
%First entry & \textbf{Second entry} & \textbf{Third entry} & \textbf{Fourth entry} \\
 \;\;\;Transition\;\;\;     & \;\;\;initial state\;\;\;    & \;\;\;final state\;\;\;    & \;\;\; $\Gamma_{e1}$($D$ meson)\;\;\; & $\Gamma_{e1}$($D_s$ meson)  \\
&$nJ^{P}$&$nJ^{P}$&($c\bar{u}$, $c\bar{d}$)\\
\hline\hline
\endfirsthead
\multicolumn{5}{c}%
{\tablename\ \thetable\ -- \textit{Continued from the previous page}} \\
\hline\hline
%\textbf{First entry} & \textbf{Second entry} & \textbf{Third entry} & \textbf{Fourth entry} \\
 \;\;\;Transition\;\;\;     & \;\;\;initial state\;\;\;    & \;\;\;final state\;\;\;    & \;\;\; $\Gamma_{e1}$($D$ meson)\;\;\; & $\Gamma_{e1}$($D_s$ meson)  \\
 &$nJ^{P}$&$nJ^{P}$&($c\bar{u}$, $c\bar{d}$)\\
\hline\hline
\endhead
\hline \multicolumn{5}{r}{\textit{Continued on the next page}} \\
\endfoot
\endlastfoot
 $0^- \rightarrow 1^+\gamma$ & $20^-$ & $11^+$ & 91.149, 3.457 & 0.4176     \\
                             &        & $21^+$ &  75.435, 1.9573  & 0.2435    \\
                             & $30^-$ & $11^+$ & 210.5 , 5.8748   & 0.3448    \\
                             &        & $21^+$ & 226.65, 5.8811   &  0.2216      \\
                             &        & $31^+$ &  432.72, 11.228  & 1.049    \\
                             &        & $41^+$ & 109.09, 2.8305  & 0.697    \\
 \hline
  $0^+ \rightarrow 1^-\gamma$& $10^+$ & $11^-$ & 728.01, 19.4031  &  0.0859   \\
                             & $20^+$ & $11^-$ & 184.37, 4.8241  &  2.0001    \\
                             &        & $21^-$ &  545.88, 14.164  &  0.2027   \\
                             &        & $31^-$ & 35.948, 0.9328   &  0.0002   \\
                             & $30^+$ & $11^-$ &  47.770, 1.2456  &  0.3441   \\
                             &        & $21^-$ & 290.88, 7.5475   & 1.6971    \\
                             &        & $31^-$ &  407.65, 10.577   & 1.6964   \\
                             &        & $41^-$ & 748.15, 19.413    & 0.5871   \\
                             &        & $51^-$ & 158.37, 4.1092  &  0.162   \\
 \hline
 $1^- \rightarrow 0^+\gamma$ & $21^-$ & $10^+$ &  43.527, 1.1612  &  1.9786   \\
                             & $31^-$ & $10^+$ & 628.92, 16.779    &  3.8272   \\
                             & $41^-$ & $10^+$ & 776.18, 20.707    &  0.1186  \\
                             &        & $20^+$ &  604.83, 16.136   &  2.8181   \\
                             & $51^-$ & $10^+$ & 49.96, 1.3328    & 0.2344   \\
                             &        & $20^+$ & 604.83, 16.136   & 2.8181   \\
  \hline
 $1^- \rightarrow 1^+\gamma$ & $21^-$ & $11^+$ & 82.384, 2.7702    & 0.4364    \\
                             &        & $21^+$ & 73.119, 1.8973    &  0.291   \\
                             & $31^-$ & $11^+$ & 497.36, 14.921    & 1.6613    \\
                             &        & $21^+$ &  463.44, 12.025  &  1.263   \\
                             & $41^-$ & $11^+$ & 1028.9, 28.573    & 0.9155    \\
                             &        & $21^+$ &  1016.9, 26.386   & 0.5979    \\
                             &        & $31^+$ &  383.55, 9.9521  & 0.6226    \\
                             &        & $41^+$ &  133.93, 3.4751   &  0.4148   \\
                             & $51^-$ & $11^+$ &  234.27, 6.4414   &   0.5734  \\
                             &        & $21^+$ &  244.65, 6.3481   & 0.3518    \\
                             &        & $31^+$ &  1126.6, 29.234   &  1.1176   \\
                             &        & $41^+$ &  585.33, 15.188   & 0.7807    \\
  \hline
    $1^+ \rightarrow 0^-\gamma$& $11^+$ & $10^-$ & 847.18, 20.415  & 3.4365  \\
                               & $21^+$ & $10^-$ &  886.97, 23.498 & 3.7954  \\
                               & $31^+$ & $10^-$ &  0.123, 0.0032  & 0.1012    \\
                               &        & $20^-$ & 443.01, 11.495  & 2.1211    \\
                               & $41^+$ & $10^-$ & 0.7203, 0.0189   &  0.0009   \\
                               &        & $20^-$ & 764.76, 19.844  &  2.7341   \\
 \hline
   $1^+ \rightarrow 1^-\gamma$& $11^+$ & $11^-$ & 638.97, 14.809  & 2.2145   \\
                              & $21^+$ & $11^-$ & 851.13, 22.510 & 2.526    \\
                              & $31^+$ & $11^-$ & 80.893, 2.1169  & 0.4596    \\
                              &        & $21^-$ & 313.77, 8.1415  & 1.8567      \\
                              &        & $31^-$ &  4.6457, 0.1205  &  0.3447   \\
                              & $41^+$ & $11^-$ & 106.76, 2.7914  &  0.5276   \\
                              &        & $21^-$ & 622.18, 16.144 &  2.2821  \\
                              &        & $31^-$ & 20.975, 0.5442 & 0.4659    \\
 \hline
  $1^+ \rightarrow 2^-\gamma$& $31^+$ & $12^-$ & 65.085, 1.6888  &   0.5529   \\
                             &        & $22^-$ &  0.2119, 0.0055 & 0.1558    \\
                             & $41^+$ & $12^-$ & 238.29, 6.1831  & 0.8464    \\
                             &        & $22^-$ & 26.278, 0.6818  & 0.2738    \\
                             & $51^+$ & $12^-$ & 1.3736, 0.0356  &  0.1362  \\
                             &        & $22^-$ & 5.0118, 0.13  & 0.1717   \\
                             &        & $32^-$ & 181.45, 4.7082  &  0.998  \\
                             &        & $42^-$ & 5.0814, 0.1318  &  0.4126  \\
                             & $61^+$ & $12^-$ & 2.6022, 0.0675  &  0.2015  \\
                             &        & $22^-$ & 0.6375, 0.0165  & 0.2359  \\
                             &        & $32^-$ & 554.7, 14.393  & 1.3456  \\
                             &        & $42^-$ & 99.485, 2.5814  & 0.5961  \\

 \hline
  $2^- \rightarrow 1^+\gamma$& $12^-$ & $11^+$ & 1324.5, 40.073  &  5.3991   \\
                             &        & $21^+$ & 1235.9, 32.0705  &  4.27   \\
                             & $22^-$ & $11^+$ & 2821.6, 81.961  &  8.0433   \\
                             &        & $21^+$ & 2692.4, 69.861  &  6.5345  \\
                             & $32^-$ & $11^+$ & 135.6, 3.736  & 0.0116   \\
                             &        & $21^+$ & 154.87, 4.0186  & 0.0318   \\
                             &        & $31^+$ & 1541.9, 40.008  &  4.8947  \\
                             &        & $41^+$ & 746.69, 19.375 & 3.7323   \\
                             & $42^-$ & $11^+$ & 167.93, 4.589  &  0.0259   \\
                             &        & $21^+$ & 194.43, 5.045  & 0.0352    \\
                             &        & $31^+$ & 3316.8, 86.062  & 6.8996    \\
                             &        & $41^+$ & 2019.9, 52.411   & 5.3872   \\
 \hline\hline
 \label{tConvE1}
\end{longtable}

%=======================================================================================================
% table Hybrid e1 transition
\LTcapwidth=\textwidth
\begin{longtable}{ccccc}
\caption{Partial widths for $E1$ decays (in KeV) of hybrid $D$ and $D_s$ mesons.}\\
\hline\hline
%First entry & \textbf{Second entry} & \textbf{Third entry} & \textbf{Fourth entry} \\
 \;\;\;Transition\;\;\;     & \;\;\;initial state\;\;\;    & \;\;\;final state\;\;\;    & \;\;\; $\Gamma_{e1}$(hybrid $D$ meson)\;\;\; & $\Gamma_{e1}$(hybrid $D_s$ meson)  \\
&$nJ^{P}$&$nJ^{P}$&($c\bar{u}$, $c\bar{d}$)\\
\hline\hline
\endfirsthead
\multicolumn{5}{c}%
{\tablename\ \thetable\ -- \textit{Continued from the previous page}} \\
\hline\hline
%\textbf{First entry} & \textbf{Second entry} & \textbf{Third entry} & \textbf{Fourth entry} \\
 \;\;\;Transition\;\;\;     & \;\;\;initial state\;\;\;    & \;\;\;final state\;\;\;    & \;\;\; $\Gamma_{e1}$(hybrid $D$ meson)\;\;\; & $\Gamma_{e1}$(hybrid $D_s$ meson)  \\
 &$nJ^{P}$&$nJ^{P}$&($c\bar{u}$, $c\bar{d}$)\\
\hline\hline
\endhead
\hline \multicolumn{5}{r}{\textit{Continued on the next page}} \\
\endfoot
\endlastfoot
 $0^{+\varepsilon} \rightarrow 1^{-\varepsilon}\gamma$ & $20^{+\varepsilon}$ & $11^{-\varepsilon}$ & 261.12, 6.7754  &  1.059   \\
                             &        & $21^{-\varepsilon}$ & 81.811, 2.1228  &  0.6407   \\
                             & $30^{+\varepsilon}$ & $11^{-\varepsilon}$ &  57.046, 1.4802 & 0.2516  \\
                             &        & $21^{-\varepsilon}$ & 60.926, 1.5809  &  0.2588      \\
                             &        & $31^{-\varepsilon}$ & 385.99, 10.016  & 1.5485    \\
                             &        & $41^{-\varepsilon}$ & 117.44, 3.0472  & 1.0873    \\
 \hline
  $0^{-\varepsilon} \rightarrow 1^{+\varepsilon}\gamma$& $10^{-\varepsilon}$ & $11^{+\varepsilon}$ & 618.71, 16.054 & 0.4149    \\
                             & $20^{-\varepsilon}$ & $11^{+\varepsilon}$ & 387.02, 10.0423  &  1.9049    \\
                             &        & $21^{+\varepsilon}$ & 264.43, 6.8614   &  0.0996   \\
                             &        & $31^{+\varepsilon}$ &  63.071, 1.6365  & 0.00002   \\
                             & $30^{-\varepsilon}$ & $11^{+\varepsilon}$ & 6.5882, 0.1709  & 0.2337    \\
                             &        & $21^{+\varepsilon}$ & 251.11, 6.5156   & 1.8328    \\
                             &        & $31^{+\varepsilon}$ & 153.29, 3.9775   & 2.48   \\
                             &        & $41^{+\varepsilon}$ & 301.26, 7.817   & 0.3068   \\
                             &        & $51^{+\varepsilon}$ & 158.79, 4.1204  &   0.0377  \\
 \hline
 $1^{+\varepsilon} \rightarrow 0^{-\varepsilon}\gamma$ & $21^{+\varepsilon}$ & $10^{-\varepsilon}$ &  199.35, 5.3184  &  1.3168   \\
                             & $31^{+\varepsilon}$ & $10^{-\varepsilon}$ &  511.52, 13.646   & 2.395    \\
                             & $41^{+\varepsilon}$ & $10^{-\varepsilon}$ & 3.2826, 0.0876    &  0.0766  \\
                             &        & $20^{-\varepsilon}$ &  231.81, 6.1844   & 2.0572    \\
                             & $51^{+\varepsilon}$ & $10^{-\varepsilon}$ & 16.243, 0.4333    & 0.2632   \\
                             &        & $20^{-\varepsilon}$ &  417.04, 11.126  & 2.48   \\
  \hline
 $1^{+\varepsilon} \rightarrow 1^{-\varepsilon}\gamma$ & $21^{+\varepsilon}$ & $11^{-\varepsilon}$ &  222.27, 5.7673   &  0.7543   \\
                             &        & $21^{-\varepsilon}$ &  85.868, 2.2281   & 0.4825    \\
                             & $31^{+\varepsilon}$ & $11^{-\varepsilon}$ & 436.29, 11.3207    &  1.7273   \\
                             &        & $21^{-\varepsilon}$ & 229.88, 5.9648   & 1.2476    \\
                             & $41^{+\varepsilon}$ & $11^{-\varepsilon}$ &  71.744, 1.8616   &  0.4743   \\
                             &        & $21^{-\varepsilon}$ &  61.15, 1.5867   &  0.3914   \\
                             &        & $31^{-\varepsilon}$ & 303.65, 7.8791   & 0.8211    \\
                             &        & $41^{-\varepsilon}$ &   115.48, 2.9963  & 0.5718    \\
                             & $51^{+\varepsilon}$ & $11^{-\varepsilon}$ &   132.82, 3.4463  &  0.1873   \\
                             &        & $21^{-\varepsilon}$ & 113.01, 2.9322    &  0.156   \\
                             &        & $31^{-\varepsilon}$ &  374.27, 9.7114   &   1.452  \\
                             &        & $41^{-\varepsilon}$ &   174.23, 4.5208  &  1.0843   \\
  \hline
    $1^{-\varepsilon} \rightarrow 0^{+\varepsilon}\gamma$& $11^{-\varepsilon}$ & $10^{+\varepsilon}$ & 420.74, 10.917  & 1.4283 \\
                               & $21^{-\varepsilon}$ & $10^{+\varepsilon}$ & 714.93, 18.551  & 1.8474  \\
                               & $31^{-\varepsilon}$ & $10^{+\varepsilon}$ & 21.638, 0.5615   &  0.000008   \\
                               &        & $20^{+\varepsilon}$ & 350.18, 9.0864  & 1.3169    \\
                               & $41^{-\varepsilon}$ & $10^{+\varepsilon}$ & 23.361, 0.6062   &  0.0033   \\
                               &        & $20^{+\varepsilon}$ &  652.03, 16.918 & 1.6334    \\
 \hline
   $1^{-\varepsilon}\rightarrow 1^{+\varepsilon}\gamma$& $11^{-\varepsilon}$ & $11^{+\varepsilon}$ & 462.52, 12.001  & 1.4309   \\
                              & $21^{-\varepsilon}$ & $11^{+\varepsilon}$ & 827.38, 21.468  &  1.8929   \\
                              & $31^{-\varepsilon}$ & $11^{+\varepsilon}$ &  163.51, 4.2426 & 0.5119    \\
                              &        & $21^{+\varepsilon}$ & 265.17, 6.8805  &  1.2207     \\
                              &        & $31^{+\varepsilon}$ &  49.942, 1.2959 &  0.473   \\
                              & $41^{-\varepsilon}$ & $11^{+\varepsilon}$ & 196.41, 5.0965  &  0.577   \\
                              &        & $21^{+\varepsilon}$ & 543.57, 14.104  & 1.5197    \\
                              &        & $31^{+\varepsilon}$ & 158.73, 4.1187 & 0.6515    \\
 \hline
  $1^{-\varepsilon} \rightarrow 2^{+\varepsilon}\gamma$& $31^{-\varepsilon}$ & $12^{+\varepsilon}$ & 251.90, 6.5361  &  1.0387  \\
                             &        & $22^{+\varepsilon}$ & 22.292, 0.5784  & 0.3843    \\
                             & $41^{-\varepsilon}$ & $12^{+\varepsilon}$ & 549.57, 14.26  &  1.3888   \\
                             &        & $22^{+\varepsilon}$ & 116.31, 3.0179  &  0.5716   \\
                             & $51^{-\varepsilon}$ & $12^{+\varepsilon}$ & 12.435, 0.3227  & 0.0632   \\
                             &        & $22^{+\varepsilon}$ & 30.026, 0.7791  & 0.1367   \\
                             &        & $32^{+\varepsilon}$ & 377.98, 9.8075  &  1.5399  \\
                             &        & $42^{+\varepsilon}$ & 38.496, 0.9989  &  0.6371  \\
                             & $61^{-\varepsilon}$ & $12^{+\varepsilon}$ & 1.709, 0.0443  & 0.0708   \\
                             &        & $22^{+\varepsilon}$ & 18.596, 0.4825  & 0.1541  \\
                             &        & $32^{+\varepsilon}$ & 860.24, 22.321  & 2.0341  \\
                             &        & $42^{+\varepsilon}$ & 211.48, 5.4873  & 0.9093  \\

 \hline
  $2^{+\varepsilon} \rightarrow 1^{-\varepsilon}\gamma$& $12^{+\varepsilon}$ & $11^{-\varepsilon}$ & 872.62, 22.642  &  3.3811 \\
                             &        & $21^{-\varepsilon}$ & 373.06, 9.6801  &  2.3478   \\
                             & $22^{+\varepsilon}$ & $11^{-\varepsilon}$ & 2390.7, 62.033  &  6.0388   \\
                             &        & $21^{-\varepsilon}$ & 1386.2, 35.968  & 4.5079   \\
                             & $32^{+\varepsilon}$ & $11^{-\varepsilon}$ & 74.05, 1.9214  &  0.0924  \\
                             &        & $21^{-\varepsilon}$ & 74.711, 1.9386  & 0.0787   \\
                             &        & $31^{-\varepsilon}$ &  736, 19.097 &  3.0566  \\
                             &        & $41^{-\varepsilon}$ & 284.58, 7.3841 &  2.3242  \\
                             & $42^{+\varepsilon}$ & $11^{-\varepsilon}$ & 57.592, 1.4944  &  0.0451   \\
                             &        & $21^{-\varepsilon}$ & 66.767, 1.7324  &  0.0386   \\
                             &        & $31^{-\varepsilon}$ & 2063.6, 53.545  &  5.1521   \\
                             &        & $41^{-\varepsilon}$ & 1143.8, 29.679   & 4.0959   \\

 \hline\hline
 \label{tHybridE1}
\end{longtable}
%=============================================================================================
%table Conventional M1 transitions
\begin{table}
\caption{Partial widths for $M1$ decays (in KeV) of conventional $D$ and $D_s$ mesons.}
\begin{tabular}{ccccc}
 \hline\hline
 \;\;\;Transition\;\;\;     & \;\;\;initial state\;\;\;    & \;\;\;final state\;\;\;    & \;\;\; $\Gamma_{m1}$($D$ meson)\;\;\; & $\Gamma_{m1}$($D_s$ meson)  \\
&$nJ^{P}$&$nJ^{P}$&($c\bar{u}$, $c\bar{d}$)\\
 \hline\hline
 $0^- \rightarrow 1^-\gamma$ & $20^-$ & $11^-$ & 11.097, 7.4747  &  2.184  \\
                             & $30^-$ & $11^-$ & 0.1487, 6.2222  &  2.3694   \\
                             &        & $21^-$ & 22.198, 10.062  &  2.9894   \\
                             &        & $31^-$ &  98.771, 1.2647  & 0.209    \\
                             & $40^-$ & $11^-$ & 0.1319, 1.566  & 0.5123    \\
                             &        & $21^-$ & 25.08, 18.489 & 5.9304    \\
                             &        & $31^-$ & 248.86, 10.857  &  0.1343   \\
                             &        & $41^-$ & 16.733, 7.8301  & 3.7441   \\
                             &        & $51^-$ &  68.052, 0.0079 & 2.0575    \\
 \hline
 $1^- \rightarrow 0^-\gamma$ & $11^-$ & $10^-$ &  23.648, 0.6064  &  0.0665   \\
                             & $21^-$ & $10^-$ & 127.55, 17.5012 &  4.0203   \\
                             &        & $20^-$ & 2.5479, 0.0644  &  0.0042   \\
                             & $31^-$ & $10^-$ & 2085, 19.985 & 7.9164   \\
                             &        & $20^-$ & 4.8418, 0.19  &   0.0919  \\
                             & $41^-$ & $10^-$ & 112.9, 17.719 &  1.622   \\
                             &        & $20^-$ & 82.281, 10.879  &  1.9332   \\
                             &        & $30^-$ &  0.6405, 0.0163 &  0.00001   \\
                             & $51^-$ & $10^-$ & 1350, 0.9538 &  5.1259   \\
                             &        & $20^-$ & 42.254, 1.3827  &  2.8697   \\
                             &        & $30^-$ & 7.1297, 0.123  &  0.004   \\
  \hline
  \hline
  \end{tabular}
   \label{tConvM1}
\end{table}
%table hybrid M1
\begin{table}
\caption{Partial widths for $M1$ decays (in KeV) of hybrid $D$ and $D_s$ mesons.}
\begin{tabular}{ccccc}
 \hline\hline
 \;\;\;Transition\;\;\;  & \;\;\;initial state\;\;\; & \;\;\;final state\;\;\; & \;\;\; $\Gamma_{m1}$(hybrid $D$ meson)\;\;\; & $\Gamma_{m1}$(hybrid $D_s$ meson)  \\
&$nJ^{P}$&$nJ^{P}$&($c\bar{u}$, $c\bar{d}$)\\
 \hline\hline
 $0^{+\varepsilon} \rightarrow 1^{+\varepsilon}\gamma$ & $20^{+\varepsilon}$ & $11^{+\varepsilon}$ & 22.206, 9.4979   & 2.1082   \\
                                                       & $30^{+\varepsilon}$ & $11^{+\varepsilon}$ & 8.4751, 10.6  &  2.2568   \\
                                                       &                     & $21^{+\varepsilon}$ & 14.759, 6.9874  &  2.1905  \\
                                                       &                     & $31^{+\varepsilon}$ &  2.6918, 1.303 &  0.5497  \\
                                                       & $40^{+\varepsilon}$ & $11^{+\varepsilon}$ & 9.428, 5.1214  &  0.7265  \\
                                                       &                     & $21^{+\varepsilon}$ & 21.874, 16.759 &  4.8863   \\
                                                       &                     & $31^{+\varepsilon}$ & 13.313, 3.7718 &  0.6848   \\
                                                       &                     & $41^{+\varepsilon}$ & 12.489, 5.5476 & 2.5828   \\
                                                       &                     & $51^{+\varepsilon}$ & 10.924, 2.235 &  1.2671  \\
 \hline
 $1^{+\varepsilon} \rightarrow 0^{+\varepsilon}\gamma$ & $11^{+\varepsilon}$ & $10^{+\varepsilon}$ & 0.2161, 0.0056 & 0.0003   \\
                                                       & $21^{+\varepsilon}$ & $10^{+\varepsilon}$ & 53.308, 7.458 &  1.1989  \\
                                                       &                     & $20^{+\varepsilon}$ & 0.4085, 0.0105 & 0.0002   \\
                                                       & $31^{+\varepsilon}$ & $10^{+\varepsilon}$ & 425.28, 17.191 & 2.2738   \\
                                                       &                     & $20^{+\varepsilon}$ & 12.745, 0.2962 & 0.0171   \\
                                                       & $41^{+\varepsilon}$ & $10^{+\varepsilon}$ & 60.381, 11.716 & 0.9809  \\
                                                       &                     & $20^{+\varepsilon}$ & 46.49, 6.0773 &  0.9054  \\
                                                       & $51^{+\varepsilon}$ & $10^{+\varepsilon}$ & 760.03, 29.874 & 2.861   \\
                                                       &                     & $20^{+\varepsilon}$ & 234.39, 9.8806 &  1.5053  \\
                                                       &                     & $30^{+\varepsilon}$ & 2.5875, 0.0632 & 0.0022   \\
  \hline
  \hline
  \end{tabular}
   \label{tHybridM1}
\end{table}
%=================================================================================================
%table comparison Radiative decays
\begin{table}
\caption{Comparison of our calculated radiative decay partial widths (in KeV) with other works. $D_1$ and $D_1^{'}$ are mixed states ($J^P$ = $1^+$ ) of charmed mesons.}
\begin{tabular}{ccccccccc}
 \hline\hline
 \;\;\;Mode\;\;\; & \;\;\;Our Results\;\;\;& \;\;\;GI Model \cite{godfrey2016}\;\;\; & \;\;\;\cite{green}\;\;\; &\;\;\;\cite{radford2009}\;\;\;&\;\;\;\cite{chen2020}&PDG \cite{PDG} \\
 \hline\hline
 Charmed strange mesons\\
 $D_s^{*\pm} \rightarrow D_s^{\pm}\gamma$ & 0.0665 & 1.03 & 2.39 & 1.12 &  & $<$ 1776.5\\
 $D_{s1}(2536)^\pm \rightarrow D_s^{\pm}\gamma$ & 3.7954 & 9.23 & 61.2 & 37.7 & 3.53 \\
 $D_{s1}(2536)^\pm \rightarrow D_s^{*\pm}\gamma$ & 2.526 & 9.61 & 9.21 & 5.74 & 4.74 \\
 Charmed mesons ($c\bar{u}$)\\
 $D^*(2007)^{0} \rightarrow D^{0}\gamma$ & 23.648 & 106 &&&& $<$ 741 \\
 $D_0^{*}(2300)^{0} \rightarrow D^{*}(2007)^{0}\gamma$ & 728.01 & 288 \\
  $D_1(2430)^{0} \rightarrow D^{0}\gamma$ & 847.18 & 640  \\
  $D_1(2420) \rightarrow D^{0}\gamma$ &  886.97 & 156 \\
 $D_1(2430)^{0} \rightarrow D^{*}(2007)^{0}\gamma$ & 638.97 & 82.8 \\
   $D_1(2420) \rightarrow D^{*}(2007)^{0}\gamma$ & 851.13 & 386 \\
  Charmed mesons ($c\bar{d}$)\\
 $D^*(2010)^{\pm}\rightarrow D^{\pm}\gamma$ & 0.6064  &  10.8 &&&&   $\approx$ 1.334 \\
 $D_0^{*} \rightarrow D^{*}(2010)^{\pm}\gamma$ & 19.4031  &  30 \\
  $D_1 \rightarrow D^{\pm}\gamma$ & 20.415  &  66  \\
  $D_1^{'}\rightarrow D^{\pm}\gamma$ & 23.498  & 16.1  \\
 $D_1 \rightarrow D^{*}(2010)^{\pm}\gamma$ & 14.809 &  8.6 \\
   $D_1^{'} \rightarrow D^{*}(2010)^{\pm}\gamma$ & 22.51  &  39.9 \\
   \hline
  \hline
  \end{tabular}
   \label{tcompRadiative}
\end{table}
\clearpage
\section{Acknowledgement}
%F. A. acknowledges HEC grant 20-15728/NRPU/R$\&$D/HEC/2021, Pakistan.
The authors would like to thank the Centre for High Energy Physics (CHEP), University of the Punjab, for providing a supportive and collaborative environment that facilitated this research.

%-------------------------

\end{document}